%% file: berhault_sips2013.tex
\pdfoutput=1
\documentclass{article}
\usepackage{spconf,amsmath,graphicx}
\usepackage{amssymb}
\usepackage{setspace}
\usepackage{textcomp}
\usepackage{tikz,pgfplots}
\usetikzlibrary{plotmarks} 
\newcommand{\Arikan}{Ar{\i}kan }

\graphicspath{{./}}

\newenvironment{alphafootnotes}
  {\par\edef\savedfootnotenumber{\number\value{footnote}}
   
   \setcounter{footnote}{0}}
  {\par\setcounter{footnote}{\savedfootnotenumber}}

\title{Partial Sums Generation Architecture\\for Successive Cancellation Decoding of Polar Codes}
\name{Guillaume Berhault, Camille Leroux, Christophe Jego, Dominique Dallet}
\address{IMS laboratory, University of Bordeaux, Institut Polytechnique de Bordeaux\\
Talence 33400, France \\e-mail: firstname.lastname@ims-bordeaux.fr}
\begin{document}
%\ninept
%
\maketitle
%
\input{berhault_sips2013_abstract}


\input{berhault_sips2013_introduction}
\input{berhault_sips2013_polar_codes}
\input{berhault_sips2013_partial_sum_computation}
\input{berhault_sips2013_results}
\input{berhault_sips2013_conclusion_and_perspectives}
%
%\section{REFERENCES}
%\label{sec:ref}

\bibliographystyle{IEEEbib}
\ninept
\bibliography{refs}

\end{document}

%% file: berhault_sips2013_abstract.tex
% % % % % % % % % % %
% Abstract
% % % % % % % % % % %

\begin{abstract}
Polar codes are a new family of error correction codes for which efficient hardware architectures have to be defined for the encoder and the decoder. Polar codes are decoded using the successive cancellation decoding algorithm that includes partial sums computations. We take advantage of the recursive structure of polar codes to introduce an efficient partial sums computation unit that can also implements the encoder. The proposed architecture is synthesized for several codelengths in 65nm ASIC technology. The area of the resulting design is reduced up to 26\% and the maximum working frequency is improved by ~25\%.
\end{abstract}

%% file: berhault_sips2013_introduction.tex
% % % % % % % % % % %
% Introduction
% % % % % % % % % % %

\section{Introduction}
\label{sec:introduction}
Polar codes \cite{arikan_channel_2008} are a new class of error correction codes. These linear block codes are proven to achieve the capacity of any symmetric memoryless channel under successive cancellation (SC) decoding \cite{sasoglu_polarization_2009}. Moreover, for a code of length $N$, encoding and decoding computational complexities are $O(N\log_2(N))$. Despite these desirable properties, polar codes require a very large code length in order to actually approach the channel capacity ($N>2^{20}$). Consequently, the practical interest of polar codes highly depends on the possibility to design efficient  encoders and decoders for large $N$ values.\\
Since polar codes invention, several hardware architectures were proposed. In \cite{arikan_channel_2008}, \Arikan suggests to use a fast Fourier transform structure to efficiently reuse computations. This first architecture requires $N\log_2(N)$ processing elements (PEs) and as many memory elements (MEs).\\
Some works then focused on reducing the number of PEs and MEs in SC decoders \cite{leroux_hardware_2011}. In \cite{leroux_hardware_2012}, a \textit{line architecture} is implemented. It only uses ($N-1$) PEs and as many MEs without affecting the decoding performance and the throughput. In \cite{leroux_semi-parallel_2012}, it is shown that the number of PEs can be further reduced ($64$ PEs) with a negligible impact on throughput. This SC decoder was fabricated in 180nm ASIC technology \cite{mishra_successive_2012}.\\
Since SC decoding has a low intrinsic parallelism, complementary works focused on increasing the throughput of SC decoders. In \cite{zhang_low-latency_2011} and \cite{zhang_low-latency_2013}, lookahead techniques are used to reduce the decoding latency while using limited extra hardware resources. 
In \cite{alamdar-yazdi_simplified_2011}, a simplification of SC decoding is proposed in order to reduce the number of computations without altering error correction performance. Extra latency reduction technique is investigated in \cite{sarkis_increasing_2013} where maximum likelihood decoding is used to further speedup the decoding process. However, these low latency decoders have not been implemented yet.\\
As shown in \cite{leroux_semi-parallel_2012} and \cite{mishra_successive_2012}, the hardware implementation of SC decoders is constrained by the partial sums computation unit which occupies a significant part of the area and limits the maximum working frequency, especially as $N$ grows. In \cite{zhang_low-latency_2013}, an alternative method to compute partial sums is proposed but was not implemented.
In this paper, we show that the partial sums computation unit can be implemented with a shift register structure, lowering hardware complexity and increasing maximum clock frequency. We also show that the proposed architecture can be used as a sequential polar code encoder.\\
% % % % % % % %
% % PLAN  % % %
% % % % % % % %
The remainder of the paper is organized as follows: in section \ref{sec:polarcodes}, the polar code construction, encoding and SC decoding processes are briefly reviewed. In the following section, the partial sums computation is introduced and a hardware implementation is proposed. Finally, in section \ref{sec:results}, this partial sums unit is compared with existing implementations in terms of area and maximum working frequency in ASIC 65nm technology.

%% file: berhault_sips2013_polar_codes.tex
% % % % % % % % % % %
% Polar codes
% % % % % % % % % % %

\section{POLAR CODES}
\label{sec:polarcodes}
\subsection{Definition and construction}
\label{ssec:subpolarconstruction}
Polar codes are linear block codes of size $N=2^n$, $n$ being a positive integer. In \cite{arikan_channel_2008}, \Arikan defined a construction based on a $2\times 2$ binary matrix, denoted as the \textit{kernel} of the code: $\kappa = \left[\begin{array}{cc}1& 0\\ 1& 1\end{array}\right] $. The generator matrix of the code is a submatrix of the $n^{th}$ Kronecker power of $\kappa$, denoted ${\kappa}^{\otimes n}$. Thus, for $n=3$ ($N=8$),

\begin{equation}
{\kappa}^{\otimes3}= \left[\begin{array}{cccccccc}1&0&0&0&0&0&0&0\\1&1&0&0&0&0&0&0\\1&0&1&0&0&0&0&0\\1&1&1&1&0&0&0&0\\1&0&0&0&1&0&0&0\\1&1&0&0&1&1&0&0\\1&0&1&0&1&0&1&0\\1&1&1&1&1&1&1&1\end{array}\right].
\label{eq:matrixg2}
\end{equation}
A polar code with dimension $K$ and codelength $N$ ($K \leq N$), is denoted as PC($N$,$K$) whose code rate is $R=\dfrac{K}{N}$. Assuming a particular successive cancellation decoding algorithm on the receiver side, the $K$ rows are selected according to the reliability of some equivalent channels. In \cite{arikan_channel_2008}, \Arikan described a method to select these $K$ rows for a binary erasure channel and a binary symetric channel. The construction was later extended to the more general binary input memoryless channels \cite{sasoglu_polarization_2009}. The reader may refer to \cite{arikan_channel_2008} for more explanations on the polarization phenomenon and polar codes definition.

\subsection{Encoding Process}
\label{ssec:subpolarcodescoding}
\begin{figure}[t]
\centering
\includegraphics[width=1\linewidth]{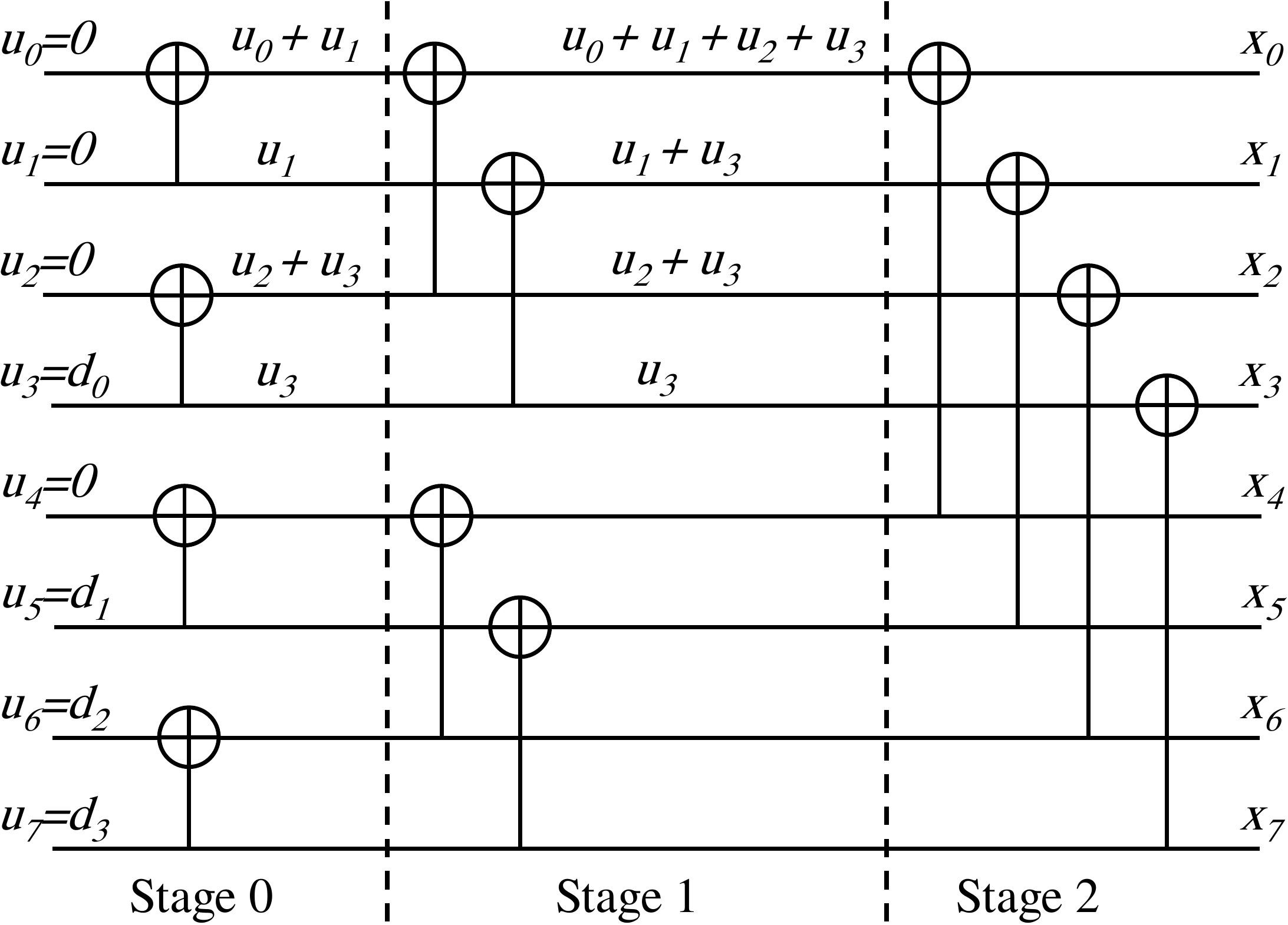}
\caption{$N=8$ polar code encoder graph.}
\label{fig:Coder_N_8_from_4-bit_coders}
\end{figure}
As any linear block code, polar code codewords are obtained by multiplying a $K$-bit information vector, $D=[d_0,d_1,...,d_{K-1}]$, with the ($K\times N$)-bit generator matrix of the code.
An alternative encoding process is to build an \textit{extended information vector} $U$ which contains the $K$ information bits and $(N-K)$ \textit{frozen bits} (all set to 0). This extended information vector is built in such a way that information bits are located on the most reliable positions corresponding to the $K$ selected rows of $\kappa^{\otimes n}$. The corresponding codeword $X$ can then be constructed by calculating $X=U\times \kappa^{\otimes n}$.\\
A polar code encoder may also be represented graphically as shown in Fig. \ref{fig:Coder_N_8_from_4-bit_coders} for $n=3$. It consists of $n$ stages of $\dfrac{N}{2}$ XORs each. The input vector $U$, on the left hand side, is propagated into the graph in order to get $X$, on the right.
%One should notice that the recursive structure of polar codes is noticeable on the encoder graph which consists of $n$ stages of $\frac{N}{2}$ XORs. As highlighted in Fig. \ref{fig:Coder_N_8_from_4-bit_coders}, a size $N$ encoder is built by applying $\frac{N}{2}$ kernel transformations to two independent sub-encoders of size $\frac{N}{2}$. Each of them is in turn constructed out of half-sized encoders. This recursive construction is applied until size 2 sub-encoders are reached.
In Fig. \ref{fig:Coder_N_8_from_4-bit_coders}, a $R=0.5$ polar code is considered; it means that half of the bits in vector $U$ are frozen bits (set to 0) while the rest of them are information bits.

\subsection{Successive Cancellation Decoding}
\label{ssec:subpolarcodesdecoding}
%In the rest of the paper, we use $v_a^b=[v_a ... v_b]$ to denote a vector $v$ indexed from $a$ to $b$, $(a,b)\in \mathbb{N}$ such as $a\leq b$.\\
\begin{figure}[t]

\begin{minipage}[b]{.48\linewidth}
  \centering
  \centerline{\includegraphics[width=4.0cm]{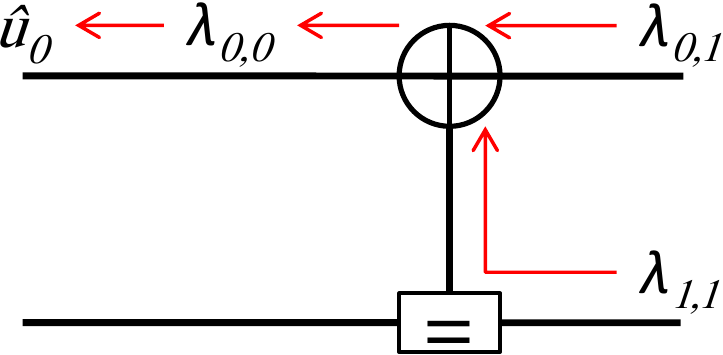}}
%  \vspace{1.5cm}
  \centerline{(a) Decoding of $\hat{u}_0$}\medskip
\end{minipage}
\hfill
\begin{minipage}[b]{0.48\linewidth}
  \centering
  \centerline{\includegraphics[width=4.0cm]{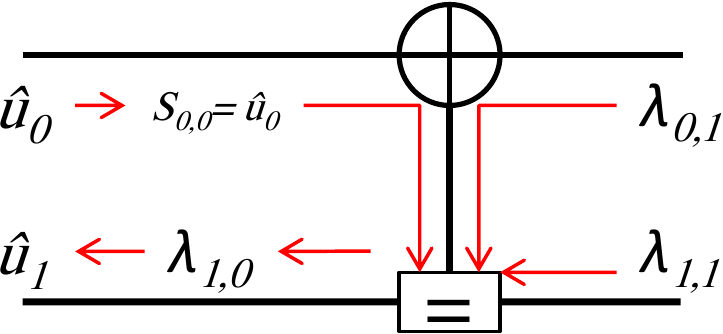}}
%  \vspace{1.5cm}
  \centerline{(b) Decoding of $\hat{u}_1$}\medskip
\end{minipage}
\caption{$N=2$ polar code decoding example.}
\label{fig:Decoder_N_2}
\end{figure}
\begin{alphafootnotes}
After being sent over the transmission channel, the noisy version $Y$ of the codeword $X$ is received. Each sample $y_i$ is converted into log likelihood ratio (LLR) format. These LLRs are denoted $\lambda_{i}$, with $0\leq i\leq N-1 $. The decoder successively estimates every bit $u_i$ based on the channel observation vector ($\lambda_0^{N-1}$)\footnote{$\lambda_0^{N-1}=[\lambda_0 ... \lambda_{N-1}]$} and the previously estimated bits ($\hat{u}_0^{i-1}$)\footnote{$\hat{u}_0^{i-1}=[\hat{u}_0 ... \hat{u}_{i-1}]$}. In order to estimate each bit $u_i$, the decoder computes the following LLR value:
\end{alphafootnotes}
\begin{equation}
\lambda _{i,0} = \log\dfrac{\Pr(y_0^{N-1},\hat{u}_0^{i-1}|u_i=0)}{\Pr(y_0^{N-1},\hat{u}_0^{i-1}|u_i=1)}.
\label{eq:llr}
\end{equation}
The estimated bit $\hat{u}_i$ is calculated based on the following rule:
\begin{equation}
\hat{u}_i=\left\{\begin{array}{ll}0 & \text{if $\lambda _{i,0}>0$}\\
1 & \text{otherwise.}\end{array}\right. 
\label{eq:u_i_decision_rule}
\end{equation}
Since the decoder knows which bits are frozen, if $u_i$ is a frozen bit, then $\hat{u}_i=0$ regardless of $\lambda_{i,0}$ value.\\
As proposed by Ar{\i}kan in \cite{arikan_channel_2008}, the factor graph representation of polar codes can be used to efficiently compute the $\lambda _{i,0}$. SC decoding can be seen as an instance of belief propagation decoding where LLRs are propagated on the factor graph of the code with a particular scheduling. In SC decoding, bits $\hat{u}_i$ are processed sequentially and the decision is then fed back into the graph for the decoding of subsequent bits. In Fig. \ref{fig:Decoder_N_2}, the factor graph of a simple $N=2$ polar code is represented. It is composed of a check node (CN or {$\oplus$}) and a variable node (VN or $\boxed{=}$). In the LLR domain, the VN function is a simple addition and the CN function uses product of transcendental functions. In the perspective of decoder implementation, the simplified versions of the VN and CN functions are used \cite{leroux_semi-parallel_2012}:\\
\begin{eqnarray}
\left\{\begin{array}{ll}
a\oplus b &= sgn(a)\times sgn(b)\times \min(|a|,|b|)\\
a\boxed{=}b &= a+b.
\end{array}\right.
\label{eq:f_g_function}
\end{eqnarray}
The decoding process of a $N=2$ polar code can be summarized as follows:
\begin{eqnarray}
\left\{\begin{array}{l}
f(\lambda_{0,1},\lambda_{1,1})= sgn(\lambda_{0,1}.\lambda_{1,1}).\min(|\lambda_{0,1}|,|\lambda_{1,1}|) \\
g(\lambda_{0,1},\lambda_{1,1},\hat{u}_0) =(-1)^{\hat{u}_0}\lambda_{0,1}+\lambda_{1,1},
\end{array}\right.
\label{eq:f_g}
\end{eqnarray}
where $\hat{u}_0$ and $\hat{u}_1$ are determined according to equation (\ref{eq:u_i_decision_rule}). The $g$ function equation shows that the decoding process of a polar code depends on LLRs propagating from right to left ($\lambda_{0,1};\lambda_{1,1}$) and also on hard decision ($\hat{u}_0$) propagating from left to right.
\begin{figure}[t]
\centering
\includegraphics[width=1\linewidth]{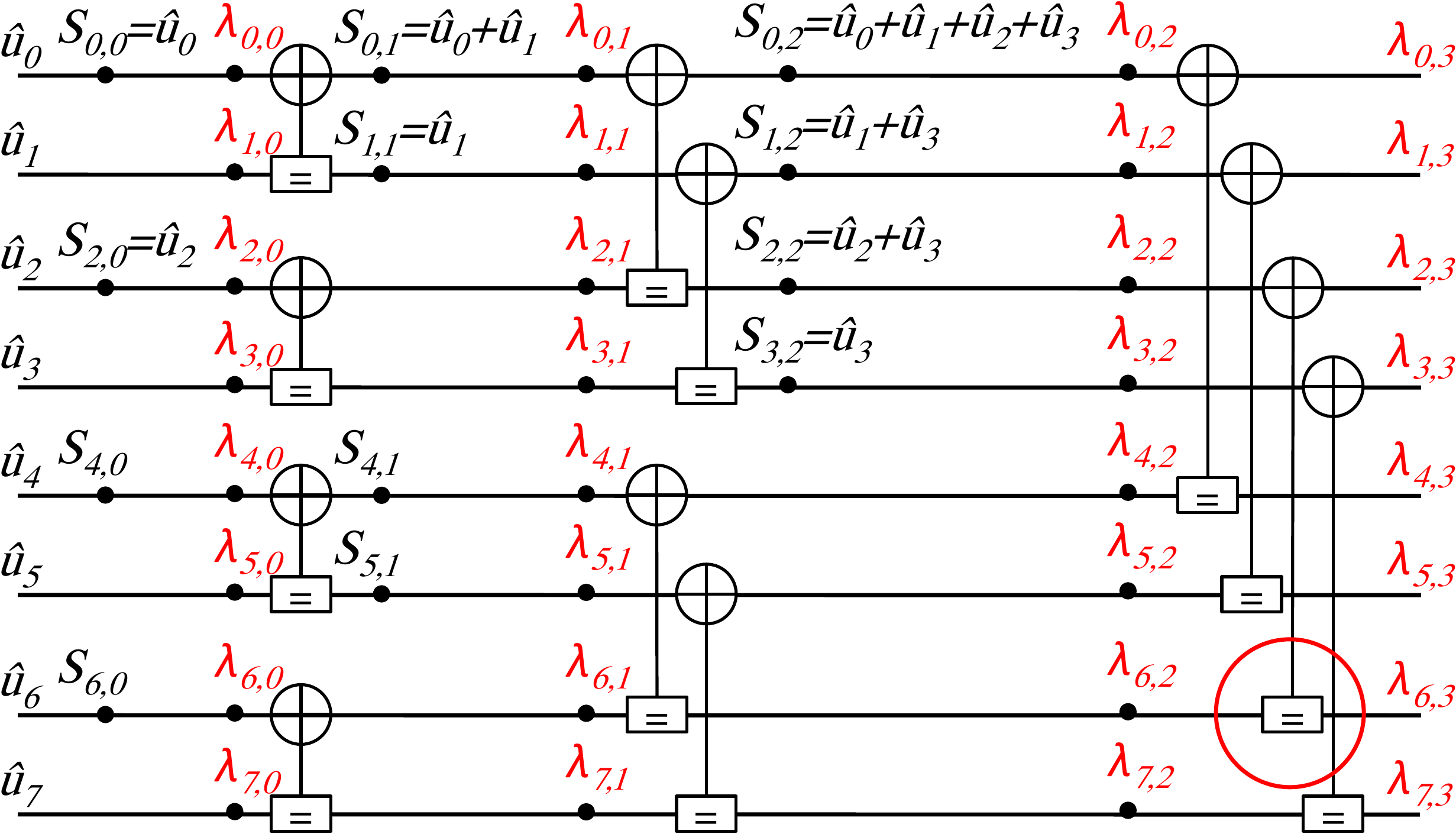}
\caption{Factor graph for $N=8$ polar code.}
\label{fig:factor_graph_N_8}
\end{figure}
The factor graph of a $N=8$ polar code is detailed in Fig. \ref{fig:factor_graph_N_8}. The decoder successively estimates the bits $\hat{u}_i$ from the computation of LLRs of the indexed edges. The LLR of edge $(i,j)$ is computed such as:
\begin{equation}
\lambda _{i,j}= \left\{\begin{array}{l l}f\left (\lambda _{i,j+1},\lambda _{i+2^j,j+1}\right) &\text{if $B(i,j)=0$}\\
g\left(\lambda _{i-2^j,j+1},\lambda _{i,j+1},S_{i-2^j,j}\right) &\text{if $B(i,j)=1$,}\end{array}\right.
\end{equation}
where $B(i,j)\equiv \dfrac{i}{2^j}\text{ mod } 2$, $0\leq i < N\text{ and } 0\leq j < n$.
$S_{i,j}$ represents the \textit{partial sum} which corresponds to the propagation of decisions back into the factor graph. For instance, in Fig. \ref{fig:factor_graph_N_8}, $S_{1,2}=\hat{u}_1+\hat{u}_3$ (modulo-2 sum).

%% file: berhault_sips2013_partial_sum_computation.tex
% % % % % % % % % % %
% Partial sum computation
% % % % % % % % % % %
\section{Partial sum computation}
\label{sec:partialsumcomputation}
When implemented in hardware, an SC decoder is composed of three main units: the \textit{processing unit} (PU), the \textit{memory unit} (MU) and the \textit{partial sums unit} (PSU). The PU consists of several processing elements (PEs) used to compute $f$ and $g$ functions. The MU stores the computed LLRs ($\lambda_{i,j}$) in register banks during the decoding process. The third unit, the PSU, computes the partial sums required by PEs to calculate the $g$ functions.\\
In \cite{leroux_semi-parallel_2012}, a semi-parallel SC decoder was synthesized for several $N$ values. Synthesis results show that the MU takes about 75\% of total decoder area while the rest of the design cost is mainly due to the PSU. In fact, the PU area becomes negligible ($<$1\%) as $N$ grows ($N>2^{13}$). As stated in \cite{leroux_semi-parallel_2012}, the MU cost can be drastically reduced by using RAM blocks instead of register banks. Consequently, the most complex part is then the PSU. Furthermore, in \cite{leroux_semi-parallel_2012} and \cite{mishra_successive_2012}, it is noticed that the critical path of the SC decoder is in the PSU and the maximum working frequency decreases as $N$ increases. Therefore, having an efficient implementation of the PSU would benefit to the SC decoder area and clock frequency.

\subsection{Existing partial sums implementations}
\label{ssec:partialsumcomputationsoa}
As depicted in Fig. \ref{fig:factor_graph_N_8}, there are $\dfrac{N}{2}\log_2(N)$ partial sums to be computed. When a bit $\hat{u}_i$ is obtained, the PSU should update all $S_{i,j}$ that include this current bit. For example, in Fig. \ref{fig:factor_graph_N_8}, when $\hat{u}_2$ is available, the partial sums \{$S_{2,0};S_{0,2};S_{2,2}$\} have to be updated by "XORing" their current values with $\hat{u}_2$. All the remaining partial sums should however keep their current values.\\
It was shown in \cite{leroux_semi-parallel_2012} that some partial sums can share the same D-Flip-Flop (DFF) thus reducing the required storage space from $\dfrac{N}{2}\log_2(N)$ to $(N-1)$ DFFs. In this work, an \textit{Indicator Function} (IF) is defined in order to indicate whether each DFF should be updated with the current $\hat{u}_i$ or not. The IF is implemented by some combinational logic that generates $(N-1)$ bits necessary to control the accumulation in the $(N-1)$ DFFs. As reported in \cite{leroux_semi-parallel_2012}, the hardware complexity of the IF-PSU increases linearly with $N$. Moreover, the number of logic gate stages in the critical path also increases with $N$. This translates into a reduction of the maximum frequency as $N$ grows.\\
In \cite{zhang_low-latency_2013}, a recursive construction of a PSU called the \emph{feedback part} (FB-PSU) is proposed. To the best of our knowledge, this architecture has not been implemented. However, from the description of the structure one can observe that the FB-PSU is composed of $(n-1)$ stages. Each one of them contains $D_l=\left(\frac{N}{2^{\log_2(N)-l+1}}+\frac{N}{2^{\log_2(N)-l+2}}\times (2^{l-2}-2)\right)$ DFFs. Therefore the total number of DFFs is $\sum _{l =2}^{n}\left(D_l \right)$. Thus $\left(\frac{N^2-4}{12}\right)$ DFFs are necessary to implement the FB-PSU along with $\left(\dfrac{N}{2}-1\right)$ XOR gates and $(N-2)$ multiplexers. Finally, the authors reported that the critical path goes through ($\log_2(N)-1$) XOR gates and ($\log_2(N)-2$) multiplexers meaning that the maximum clock frequency is affected as $N$ grows.\\
In this paper, a reduced complexity PSU architecture is described. The critical path includes few logic gates which enables to reach a high working frequency. Moreover, the proposed structure can also be used as a sequential polar code encoder.

\subsection{Shift-register-based partial sums computation unit (SR-PSU)}
\label{ssec:partialsumcomputationproposedsolution}
\begin{figure*}[t]
\centering
\includegraphics[width=0.89\linewidth]{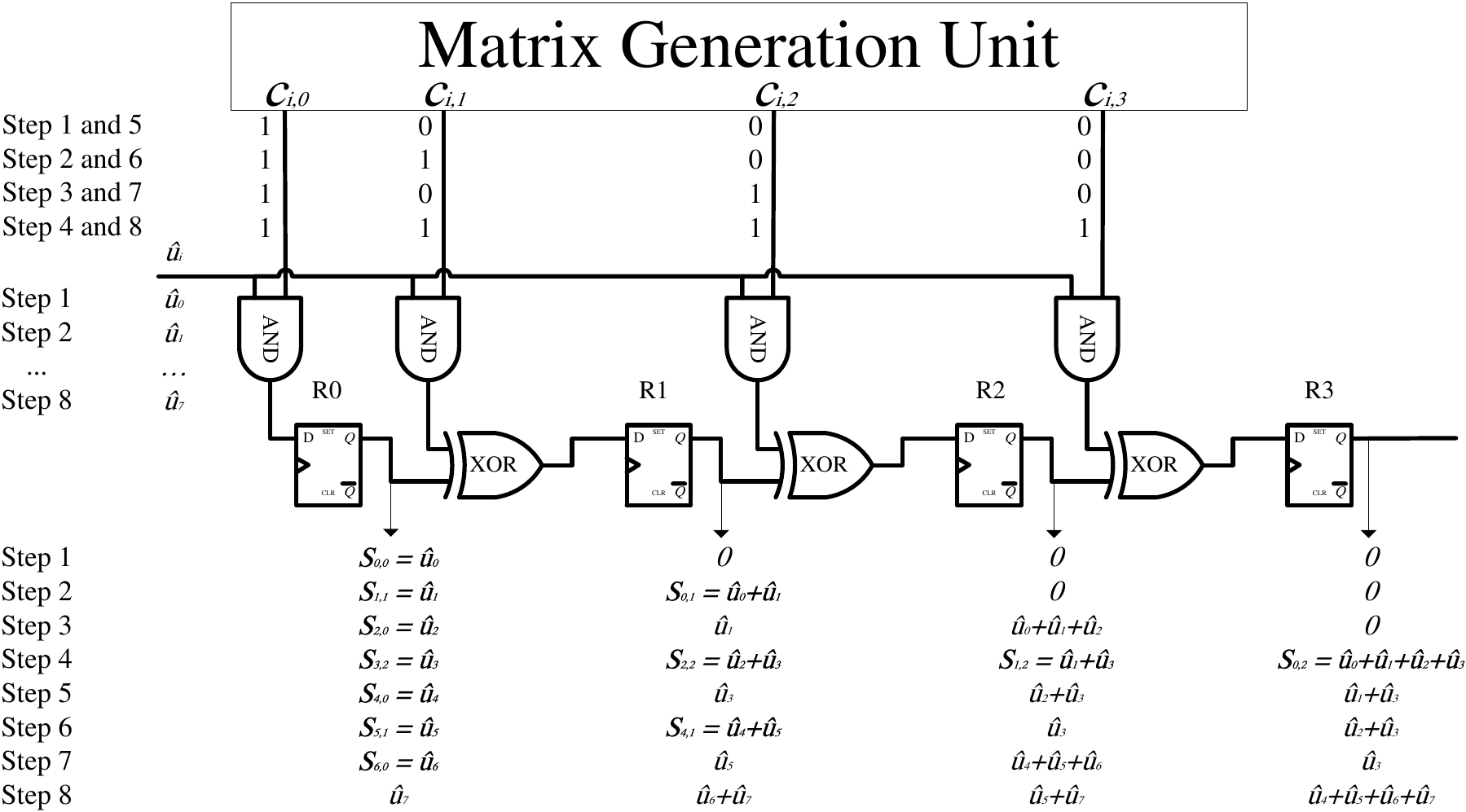}
\caption{SR-PSU example for $N=8$.}
\label{fig:ps_architecture_N_8_example}
\end{figure*}
During SC decoding process, a maximum of $\frac{N}{2}$ $g$ functions can be performed in parallel. Therefore, we propose to store the partial sums in an $\frac{N}{2}$ bit register. In addition to the $\frac{N}{2}$ DFFs, it consists of $\frac{N}{2}$ XOR gates, $\frac{N}{2}$ AND gates and a \textit{matrix generation unit} whose structure is detailed in the next section. Fig. \ref{fig:ps_architecture_N_8_example} details the proposed partial sums computation structure for $N=8$.\\
In this architecture, each DFF $R_k$ receives the value of $R_{k-1}$ which is first added (XOR) with the current decoded bit $\hat{u}_i$ if the control bit $c_{i,k}=1$. This architecture can be devised for any code length $N$ according to the following rule:
\begin{equation}
\left\{\begin{array}{lclr} R_{0}& \Leftarrow & \hat{u}_i \text{ AND } c_{i,0}&\\
R_k& \Leftarrow & R_{k-1} \text{ XOR } (\hat{u}_i \text{ AND } c_{i,k}) & \text{if $k>0$}\end{array}\right. 
\label{eq:Ri_shifting_controlled}
\end{equation}
In Fig. \ref{fig:ps_architecture_N_8_example}, the current value of the shift-register is given for each step. A step corresponds to the generation of a new bit $\hat{u}_i$. This structure generates all the required partial sums (bold values in Fig. \ref{fig:ps_architecture_N_8_example}). This shift-register structure was selected so that all partial sums required by a PE are all generated in the same DFF. It means that this SR-PSU can be included in a line SC decoder by simply connecting the $\frac{N}{2}$ PEs to a single DFF $R_k$. This avoids any extra multiplexing logic to route the partial sums to the PEs. This PSU can then be used as such for a tree or a line SC decoder. For the semi-parallel architecture, some multiplexing logic is required, exactly like in \cite{leroux_semi-parallel_2012}.\\
Although this architecture produces partial sums, it also encodes an $\dfrac{N}{2}$-bit vector. In Fig. \ref{fig:ps_architecture_N_8_example}, at step 4, each DFF $R_k$ contains the bit $x_{\frac{N}{2}-1-k}$ such that $X=U\times \kappa^{\otimes 2}$. A polar code encoder for a code length of $N$ can then be devised with $N$ DFFs, $N$ AND gates, $(N-1)$ XORs and a matrix generation unit. After bits $u_i$ are sequentially shifted during $N$ clock cycles, the codeword $X$ is contained in the register. To the best of our knowledge, this is the first reported sequential encoder for polar codes.

\subsection{Matrix generation unit}
\label{ssec:controlarchps}
\begin{figure}[t]
\centering
\includegraphics[width=1\linewidth]{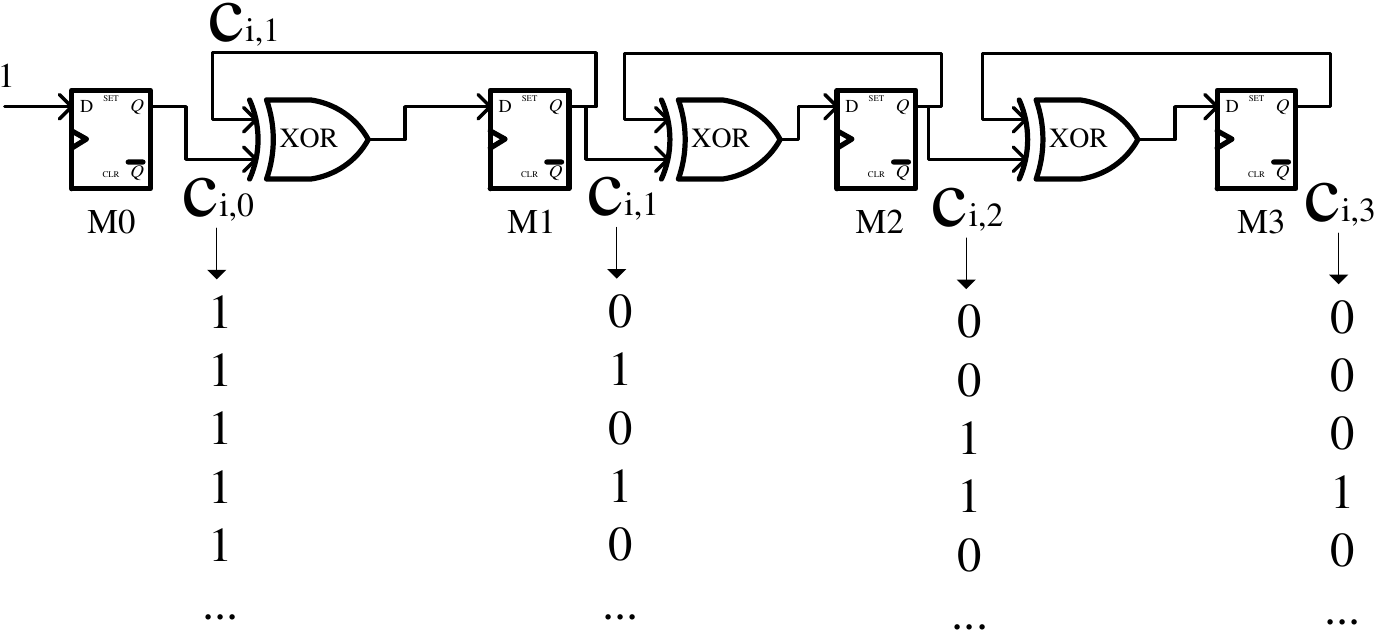}
\caption{Matrix generation unit.}
\label{fig:control_bit_gen_N_8}
\end{figure}

Let us define the control matrix as $C = \left [ \begin{array}{c} \kappa^{\otimes n-1}\\\kappa^{\otimes n-1} \end{array} \right ]$ whose element $c_{i,k}$ is the $k^{th}$ control bit generated at step $i$. In order to implement the generation of the matrix, a naive approach is to store $C$ in a ROM of size $N \times \frac{N}{2}$. This would be very complex to implement especially for large $N$ values. For this reason, a solution based on a linear feedback shift register (LFSR) is proposed to generate the matrix $\kappa^{\otimes n-1}$.
As shown in Fig. \ref{fig:ps_architecture_N_8_example}, the matrix generation unit has to produce the rows of $C$ sequentially. We propose to use an LFSR of size $\dfrac{N}{2}$ to generate this sequence. In such a structure, the state of the LFSR at step $i$ corresponds to the $i^{th}$ row of $C$. By observing the matrix C, one can verify that:
\begin{equation}
\left\{
\begin{array}{lclr}
c_{i,0} & = & 1 & 0\leq i\leq N-1\\
\\
c_{i+1,k} & = & c_{i,k-1} \text{ XOR } c_{i,k} & 0\leq i < N-1\\
& & & 1\leq k\leq \dfrac{N}{2}-1\\
\end{array} 
\right.
\label{eq:matrix_property}
\end{equation}
Let us denote the DFF that generates $c_{i,k}$ as $M_k$. From Equation \ref{eq:matrix_property}, one can deduce that matrix $C$ is generated by applying the following mapping rule to the LFSR:
\begin{equation}
\left\{
\begin{array}{llr}
M_0 & \Leftarrow 1&\\
M_k & \Leftarrow M_k \text{ XOR } M_{k-1} &\text{if $k > 0$}
\end{array}
\right.
\label{eq:matrix_property_m}
\end{equation}
This is illustrated in Fig. \ref{fig:control_bit_gen_N_8} where the matrix $\kappa^{\otimes 2}$ is generated twice which correspond to the matrix $C$ for $N=8$.

%% file: berhault_sips2013_results.tex
% % % % % % % % % % %
% Results
% % % % % % % % % % %
\section{Implementation results and comparisons}
\label{sec:results}
To the best of our knowledge IF-PSU \cite{leroux_hardware_2012} is the only implemented partial sums unit in the literature. In this section, SR-PSU logic synthesis results are provided and compared with IF-PSU in terms of area and maximum working frequency. All syntheses were performed using the low power ST Microelectronics 65nm standard cell library with supply voltage 1.0V and nominal temperature 25\textcelsius. The FB-PSU architecture, introduced in \cite{zhang_low-latency_2013}, was not implemented. However, the authors give some insights on the hardware complexity and the critical path depth. These architectural estimations are used to perform the comparison with the proposed SR-PSU.

\subsection{Functional verification and implementation methodology}
\label{ssec:verif_implementation}
\begin{alphafootnotes}
The PSU architecture introduced in section \ref{sec:partialsumcomputation} was included in a tree SC decoder described in VHDL. A set of tree SC decoders was generated for different codelengths ($2^3<N<2^{10}$)\footnote{Larger codelengths were not verified due to very long post-synthesis simulation runtime}. The resulting designs were synthesized in ASIC technology and validated by post-synthesis simulations with more than 2500 test vectors. These test vectors were obtained by a software SC decoder reference simulator that includes an AWGN channel model. Noisy codewords were generated at 7 different SNR values ranging from 0dB to 3dB. The behavior of IF-PSU and SR-PSU can be considered identical since in both architectures, $\hat{u}_i$ is shifted in sequentially and partial sums are generated in parallel. In order to perform a fair comparison, IF-FSU and SR-FSU were synthesized with the same technology and constraints.
\end{alphafootnotes}

\subsection{Hardware complexity}
\label{ssec:area}
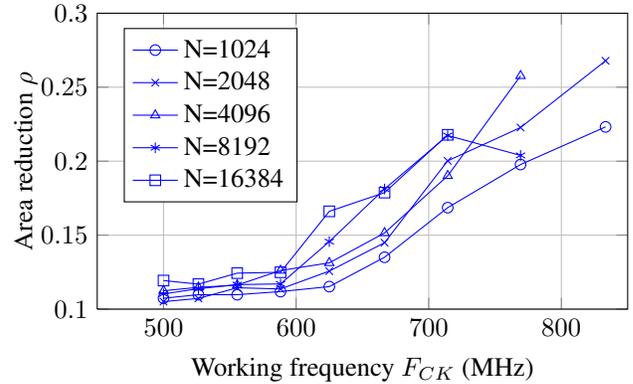
\begin{figure}[t]
\input{Area.tikz}
\caption{Area reduction \textit{vs} working frequency.}
\label{fig:Area}
\end{figure}
In order to have a fair comparison between the two architectures in terms of area, the clock frequency was set to a relatively low value ($F_{CK}=500\text{MHz}$) for the ST 65nm technology, so that the synthesis tool does not insert extra buffers or large cells in the design. Thus, the reported area corresponds to the minimum achievable hardware complexity. IF-PSU and SR-PSU designs were synthesized for different codelengths, $2^{10}\leq N \leq2^{14}$. For both architectures, the reported area is linear with $N$. In average, for all investigated codelengths, the SR-PSU architecture is 12\% smaller than the corresponding IF-FSU architecture.\\
The same designs were synthesized for higher frequency values. In such a case, the synthesis tool optimizes the design resulting in an increased area. Fig. \ref{fig:Area} shows the area reduction provided by the SR-PSU architecture for a range of target clock frequencies. The area reduction is calculated as $\rho=(\text{Area}_{\text{IF-PSU}}-\text{Area}_{\text{SR-PSU}})/\text{Area}_{\text{IF-PSU}}$. The area reduction significantly increases with the clock frequency and the codelength. It reaches 26\% for $N=4096$ and $F_{CK}=769\text{MHz}$. For each curve, the highest reported frequency corresponds to the maximum achievable frequency for the IF-PSU. In the case of SR-PSU, frequency can be pushed further as seen in the following section. The different curves show that a significant area reduction is achieved.\\
The hardware complexity estimations of FB-PSU are reported in Table \ref{tab:ressource usage} in terms of DFFs, XORs and MUXes as a function of $N$. These values are compared with the SR-PSU hardware complexity. The estimated gate count is obtained by replacing each logic operator (DFF, XOR, ...) with its equivalent NAND gate count provided by the datasheet of the ST 65nm ASIC library. The IF-PSU gate count is estimated after the area synthesis results.
%NAND gate -> 2.08 µm²
%MUX gate ->  4.68
%DFF gate -> 10
%AND gate -> 2
%XOR gate -> 4.08
%The equivalent is given as NAND gate
\begin{table}
\begin{center}
\begin{tabular}{c|ccc}
 & FB-PSU & SR-PSU & IF-PSU \\ 
\hline DFF & $\frac{N^2-4}{12}$ & $N$&no data \\ 
\hline XOR & $\frac{N}{2}-1$ & $N-2$&no data \\ 
\hline MUX & $N-2$ & - &no data\\ 
\hline AND & - &  $\frac{N}{2}-1$&no data \\ 
\hline  \hline \\ % Estimated gate & & \\
\textbf{NAND equivalent} & \boldmath$\dfrac{5N^2}{12}+3N$ & \boldmath$\dfrac{15}{2}N$ & \boldmath$\dfrac{17}{2}N$\\ 
\end{tabular} 
\end{center}
\caption{Estimated NAND gate count comparison.}
\label{tab:ressource usage}
\end{table}
For a small code length, $N=1024$, FB-PSU gate count is roughly 440,000 gates while the SR-PSU architecture only consists of 7,680 gates for the same codelength. In fact, the very high complexity of the FB-PSU is due to the number of DFFs that grows with $N^2$, making this architecture non realistic for large codelengths. Since these estimations are carried out at a low frequency ($500$MHz, no optimization), the SR-PSU and IF-PSU seem to be equivalent. Nevertheless, as seen in Fig. \ref{fig:Area}, the IF-PSU area increases more than the SR-PSU.
\subsection{Working frequency}
\label{ssec:resultsFrequency}
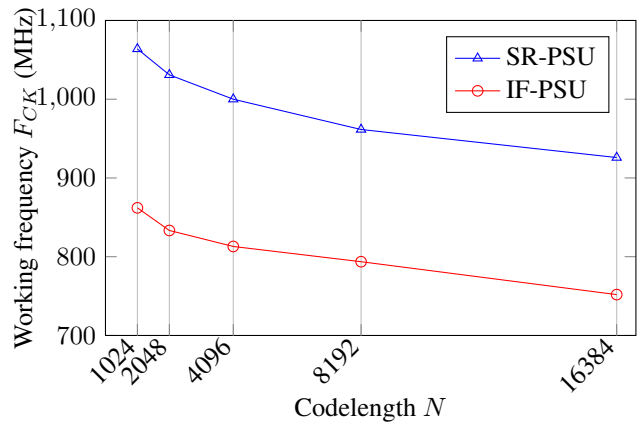
\begin{figure}[t]
\input{Fmax_N.tikz}
\caption{Maximum working frequency \textit{vs} codelength.}
\label{fig:Fmax_N}
\end{figure}

In order to estimate the maximum working frequency of IF-PSU and SR-PSU architectures, all designs were synthesized under increasing timing constraints until the synthesis tool fails at meeting the timing constraint. Fig. \ref{fig:Fmax_N} shows that for both architectures, the maximum frequency decreases with $N$. For each $N$ value, the SR-PSU reaches a higher maximum frequency than the IF-PSU. This confirms that the proposed SR-PSU has a shorter critical path than the IF-PSU. A detailed analysis of the synthesized SR-PSU designs show that the critical path starts from input $\hat{u}_i$ and ends on each of the $N$ DFFs. This means that input $\hat{u}_i$ 
drives $N$ AND gates resulting in a high fanout net. The synthesis tool has to insert extra buffers on this path in order to meet the timing constraint. This explains that in spite of the constant logic gate number included in the critical path, the SR-PSU maximum working frequency decreases with $N$.\\
As mentioned in section \ref{ssec:partialsumcomputationsoa}, FB-PSU critical path is composed of ($\log_2(N)-1$) XOR gates and ($\log_2(N)-2$) MUXes while the SR-PSU consists of only $1$ AND gate and $1$ XOR gate. This should result in a lower clock frequency. In the FB-PSU, despite the longer critical path, the input $\hat{u}_i$ has a smaller fanout which may result in a reasonable frequency. However the high hardware complexity of the FB-PSU design makes the routing phase critical and consequently affects the maximum working frequency.

%% file: Area.tikz
% This file was created by matlab2tikz v0.3.3.
% Copyright (c) 2008--2013, Nico Schlömer <nico.schloemer@gmail.com>
% All rights reserved.
% 
% The latest updates can be retrieved from
%   http://www.mathworks.com/matlabcentral/fileexchange/22022-matlab2tikz
% where you can also make suggestions and rate matlab2tikz.
% 
% 
% 
\begin{tikzpicture}

\begin{axis}[%
width=0.82\linewidth,
height=1.55in,
unbounded coords=jump,
scale only axis,
xmin=450,
xmax=850,
xlabel={Working frequency $F_{CK}$ (MHz)},
xmajorgrids,
ymin=0.1,
ymax=0.3,
ylabel style={yshift=-8pt},
ylabel={Area reduction $\rho$},
ymajorgrids,
legend style={draw=black,fill=white,legend cell align=left, at={(0.05,0.95)}, anchor=north west}
]
\addplot [
color=blue,
solid,
mark=o,
mark options={solid}
]
table[row sep=crcr]{
500 0.107371531768183\\
526.315789473684 0.109746612632527\\
555.555555555555 0.109746612632527\\
588.235294117647 0.111855270143341\\
625 0.115215622457282\\
666.666666666667 0.135049126854688\\
714.285714285714 0.168584865980958\\
769.230769230769 0.197727188872081\\
833.333333333333 0.223161199812947\\
};
\addlegendentry{N=1024};

\addplot [
color=blue,
solid,
mark=x,
mark options={solid}
]
table[row sep=crcr]{
500 0.105081209344189\\
526.315789473684 0.107217925433137\\
555.555555555555 0.114428565645116\\
588.235294117647 0.113639126382755\\
625 0.125684589246452\\
666.666666666667 0.144909407188789\\
714.285714285714 0.200215718176999\\
769.230769230769 0.222742662666877\\
833.333333333333 0.267850771736601\\
};
\addlegendentry{N=2048};

\addplot [
color=blue,
solid,
mark=triangle,
mark options={solid}
]
table[row sep=crcr]{
500 0.112417623704065\\
526.315789473684 0.114718899703252\\
555.555555555555 0.116240508760755\\
588.235294117647 0.126100996770743\\
625 0.131260125028574\\
666.666666666667 0.151476912025937\\
714.285714285714 0.190106458875742\\
769.230769230769 0.257670324451756\\
};
\addlegendentry{N=4096};

\addplot [
color=blue,
solid,
mark=asterisk,
mark options={solid}
]
table[row sep=crcr]{
500 0.110274999235264\\
526.315789473684 0.113767589719631\\
555.555555555555 0.116567025913345\\
588.235294117647 0.117038716075368\\
625 0.145640669137338\\
666.666666666667 0.181200691895541\\
714.285714285714 0.21735314007282\\
769.230769230769 0.203895243675643\\
};
\addlegendentry{N=8192};

\addplot [
color=blue,
solid,
mark=square,
mark options={solid}
]
table[row sep=crcr]{
500 0.119306491305926\\
526.315789473684 0.116883773615438\\
555.555555555555 0.124314919094013\\
588.235294117647 0.124912933017303\\
625 0.166097623868281\\
666.666666666667 0.178751719354346\\
714.285714285714 0.217827682464009\\
};
\addlegendentry{N=16384};
\end{axis}
\end{tikzpicture}%

%% file: Fmax_N.tikz
% This file was created by matlab2tikz v0.3.3.
% Copyright (c) 2008--2013, Nico Schlömer <nico.schloemer@gmail.com>
% All rights reserved.
% 
% The latest updates can be retrieved from
%   http://www.mathworks.com/matlabcentral/fileexchange/22022-matlab2tikz
% where you can also make suggestions and rate matlab2tikz.
% 
% 
% 
\begin{tikzpicture}

\begin{axis}[%
width=0.82\linewidth,
height=1.65in,
scale only axis,
xmin=0,
xmax=17000,
xtick=\empty,
%remove the scientific exponenet on the x axis
scaled x ticks = false,
% force the x label format to non scientific
x tick label style={/pgf/number format/fixed},
extra x ticks ={1024,2048,4096,8192,16384},
extra x tick style={grid=major,	tick label style={rotate=45,anchor=east}},
extra x tick labels={1024,2048,4096,8192,16384},
xlabel={Codelength $N$},
% shift the axis label
xlabel style={yshift=-5pt},
ymin=700,
ymax=1100,
ylabel={Working frequency $F_{CK}$ (MHz)},
ylabel style={yshift=-5pt},
legend style={draw=black,fill=white,legend cell align=left, at={(0.95,0.95)}, anchor=north east}
]
\addplot [
color=blue,
solid,
mark=triangle,
mark options={solid}
]
table[row sep=crcr]{
1024 1063.82978723404\\
2048 1030.92783505155\\
4096 1000\\
8192 961.538461538462\\
16384 925.925925\\
};
\addlegendentry{SR-PSU};

\addplot [
color=red,
solid,
mark=o,
mark options={solid}
]
table[row sep=crcr]{
1024 862.068965517241\\
2048 833.333333333333\\
4096 813.008130081301\\
8192 793.650793650794\\
16384 751.87969924812\\
};
\addlegendentry{IF-PSU};

\end{axis}
\end{tikzpicture}%

%% file: berhault_sips2013_conclusion_and_perspectives.tex
% % % % % % % % % % %
% Conclusion & perspectives
% % % % % % % % % % %

\section{Conclusion and perspectives}
\label{sec:conclu}
The hardware implementation of polar code decoders is a decisive step towards their potential inclusion in future digital telecommunication standards. Recent works paved the way for the definition of efficient decoder architectures. In current state of the art polar code successive cancellation decoders, the limiting element is the partial sums computation unit. In this paper, we propose a new partial sums computation architecture with improved working frequency and reduced hardware complexity. The resulting design was verified and synthesized using ASIC 65nm technology and favorably compares with state of the art partial sums units. Moreover, we also showed how this structure can be used as a sequential polar encoder. 
This new computation method opens the way for several interesting research topics such as the extension of this architecture to higher kernels or the enhancement of parallelism in this structure.